\documentclass[prb,twocolumn,showpacs,amsmath,amssymb,floatfix]{revtex4}
\usepackage{color,graphics,epsfig,rotating}
\usepackage{graphicx}% Include figure files
\usepackage{dcolumn}% Align table columns on decimal point
\usepackage{bm}% bold math
\usepackage{subfigure}

\usepackage{mathrsfs}

\def\beq{\begin{equation}}
\def\eeq{\end{equation}}

\newcommand{\zge}{z_{\textrm{Ge}}}

\begin{document}

\title{Unconventional sign-changing superconductivity near quantum
  criticality in YFe$_2$Ge$_2$}

\author{Alaska Subedi} \affiliation{Centre de Physique Th\'eorique,
  \'Ecole Polytechnique, CNRS, 91128 Palaiseau Cedex, France}

\date{\today}

\begin{abstract}
  I present the results of first principles calculations of the
  electronic structure and magnetic interactions for the recently
  discovered superconductor YFe$_2$Ge$_2$ and use them to identify the
  nature of superconductivity and quantum criticality in this
  compound. I find that the Fe $3d$ derived states near the Fermi
  level show a rich structure with the presence of both linearly
  dispersive and heavy bands. The Fermi surface exhibits nesting
  between hole and electron sheets that manifests as a peak in the
  susceptibility at $(1/2,1/2)$. I propose that the superconductivity
  in this compound is mediated by antiferromagnetic spin fluctuations
  associated with this peak resulting in a $s_\pm$ state similar to
  the previously discovered iron-based superconductors. I also find
  that various magnetic orderings are almost degenerate in energy,
  which indicates that the proximity to quantum criticality is due to
  competing magnetic interactions.
\end{abstract}


\maketitle

% \section{Introduction}

Unconventional superconductivity and quantum criticality are two of
the most intriguing phenomena observed in physics. The underlying
mechanisms and the properties exhibited by the systems in which these
two phenomena occur has not been fully elucidated because
unconventional superconductors and materials at quantum critical point
are so rare. The dearth of realizable examples has also held back the
study of the relationship and interplay between unconventional
superconductivity and quantum criticality, if there are any.

Therefore, the recent report of non-Fermi liquid behavior and
superconductivity in YFe$_2$Ge$_2$ by Zou \textit{et al.}\ is of great
interest despite a low superconducting $T_c$ of $\sim$1.8
K.\cite{zou13} This material is also interesting because it shares
some important features with the previously discovered iron-based
high-temperature superconductors. Like the other iron-based
superconductors, its structural motif is a square plane of Fe that is
tetrahedrally coordinated, in this case, by Ge. This Fe$_2$Ge$_2$
layer is stacked along the $z$ axis with an alternating layer of Y
ions.  The resulting body-centered tetragonal structure ($I4/mmm$) of
this compound is the same as that of the `122' family of the
iron-based superconductors.
%, is shown in Fig.~\ref{fig:yfg-struct}.

The nearest neighbor Fe--Ge and Fe--Fe distances of 2.393 and 2.801
\AA, respectively, in this compound\cite{vent96} are similar to the
Fe--As and Fe--Fe distances of 2.403 and 2.802 \AA, respectively,
found in BaFe$_2$As$_2$.\cite{rott08} This raises the possibility
that the direct Fe--Fe hopping is important to the physics of this
material, which is the case for the previously discovered iron-based
superconductors.\cite{sing08a}

Furthermore, Zou \textit{et al.}\ report that the superconductivity in
this compound exists in the vicinity of a quantum critical point that
is possibly associated with antiferromagnetic spin
fluctuations.\cite{zou13} A related isoelectronic compound
LuFe$_2$Ge$_2$ that occurs in the same crystal structure exhibits
antiferromagnetic spin density wave order below 9
K,\cite{avil04,fers06} and the magnetic transition is continuously
suppressed in Lu$_{1-x}$Y$_x$Fe$_2$Ge$_2$ series as Y content is
increased, with the quantum critical point lying near the composition
Lu$_{0.81}$Y$_{0.19}$Fe$_2$Ge$_2$.\cite{ran11} The proximity of
YFe$_2$Ge$_2$ to quantum criticality is observed in the non-Fermi
liquid behavior of the specific-heat capacity and resistivity.  Zou
\textit{et al.}\ find that the unusually high Sommerfeld coefficient
with a value of $C/T \simeq 90$ mJ/mol K$^2$ at 2 K further increases
to a value of $\sim$100 mJ/mol K$^2$ as the temperature is lowered,
although the experimental data is not detailed enough to distinguish
between a logarithmic and a square root increase. They also find that
the resistivity shows a behavior $\rho \propto T^{3/2}$ up to a
temperature of 10 K.

% \begin{figure}
%   \includegraphics[width=0.5\columnwidth]{YFG-fig-1}
%   \caption{(Color online) Crystal structure of body-centered
%     tetragonal YFe$_2$Ge$_2$.}
%   \label{fig:yfg-struct}
% \end{figure}

% It has been proposed that antiferromagnetic spin fluctuations are
% responsible for unconventional superconductivity in the iron-based
% compounds. It has also been suggested that these are also near a
% magnetic quantum critical point, and there are some experimental
% evidences for non-Fermi liquid behavior in certain members of this
% compound. However, the relationship and interplay between
% unconventional superconductivity and quantum criticality, if any, has
% not been elucidated clearly. Therefore, the presence of
% superconductivity and quantum critical behavior in YFe$_2$Ge$_2$
% raises an important prospect for the study of such phenomena.

In this paper, I use the results of first principles calculations to
discuss the interplay between superconductivity and quantum
criticality in YFe$_2$Ge$_2$ in terms of its electronic structure and
competing magnetic interactions. I find that the fermiology in this
compound is dominated by Fe $3d$ states with the presence of both
heavy and linearly dispersive bands near the Fermi level. The Fermi
surface consists of five sheets. There is an open tetragonal electron
cylinder around $X = (1/2, 1/2, 0)$. A large three dimensional closed
sheet that is shaped like a shell of a clam is situated around $Z =
(0, 0, 1/2) = (1, 0, 0)$. This sheet encloses a cylindrical and two
almost spherical hole sheets. The tetragonal cylinder sheet around $X$
nests with the spherical and the cylindrical sheets around $Z$, which
manifests as a peak at $(1/2,1/2)$ in the bare susceptiblity. I
propose that the superconductivity in this compound is mediated by
antiferromagnetic spin fluctuations associated with this peak, and the
resulting superconductivity has a sign-changing $s_\pm$ symmetry with
opposite signs on the nested sheets around $X$ and $Z$. This
superconductivity is similar to the one proposed for previously
discovered iron-based superconductors.\cite{mazi08,kuro08}
Furthermore, I find that there are competing magnetic interactions in
this compound, and the quantum criticality is due to the fluctuations
associated with these magnetic interactions.

% COMPUTATIONAL DETAILS

The results presented here were obtained within the local density
approximation (LDA) using the general full-potential linearized
augmented planewave method as implemented in the WIEN2k software
package.\cite{wien2k} Muffin-tin radii of 2.4, 2.2, and 2.1 a.u.\ for
Y, Fe, and Ge, respectively, were used. A $24 \times 24 \times 24$
$k$-point grid was used to perform the Brillouin zone integration in
the self-consistent calculations. An equivalently sized or larger grid
was used for supercell calcualtions. Some magnetic calculations were
also checked with the ELK software package.\cite{elk}
I used the experimental parameters ($a$ = 3.9617 and $c$ = 10.421
\AA),\cite{vent96} but employed the internal coordinate for Ge $\zge$
= 0.3661 obtained via non-spin-polarized energy minimization. The
calculated value for $\zge$ is different from the experimentally
determined value of $\zge$ = 0.3789. The difference in the Ge height
between the calculated and experimental structures is 0.13 \AA, which
is larger than the typical LDA error in predicting the crystal
structure. Such a discrepancy is also found in the iron-based
superconductors.\cite{sing08a} This may suggest that YFe$_2$Ge$_2$
shares some of the underlying physics with the previously discovered
iron-based superconductors.

\begin{figure} %% [tbp]
  \includegraphics[width=\columnwidth]{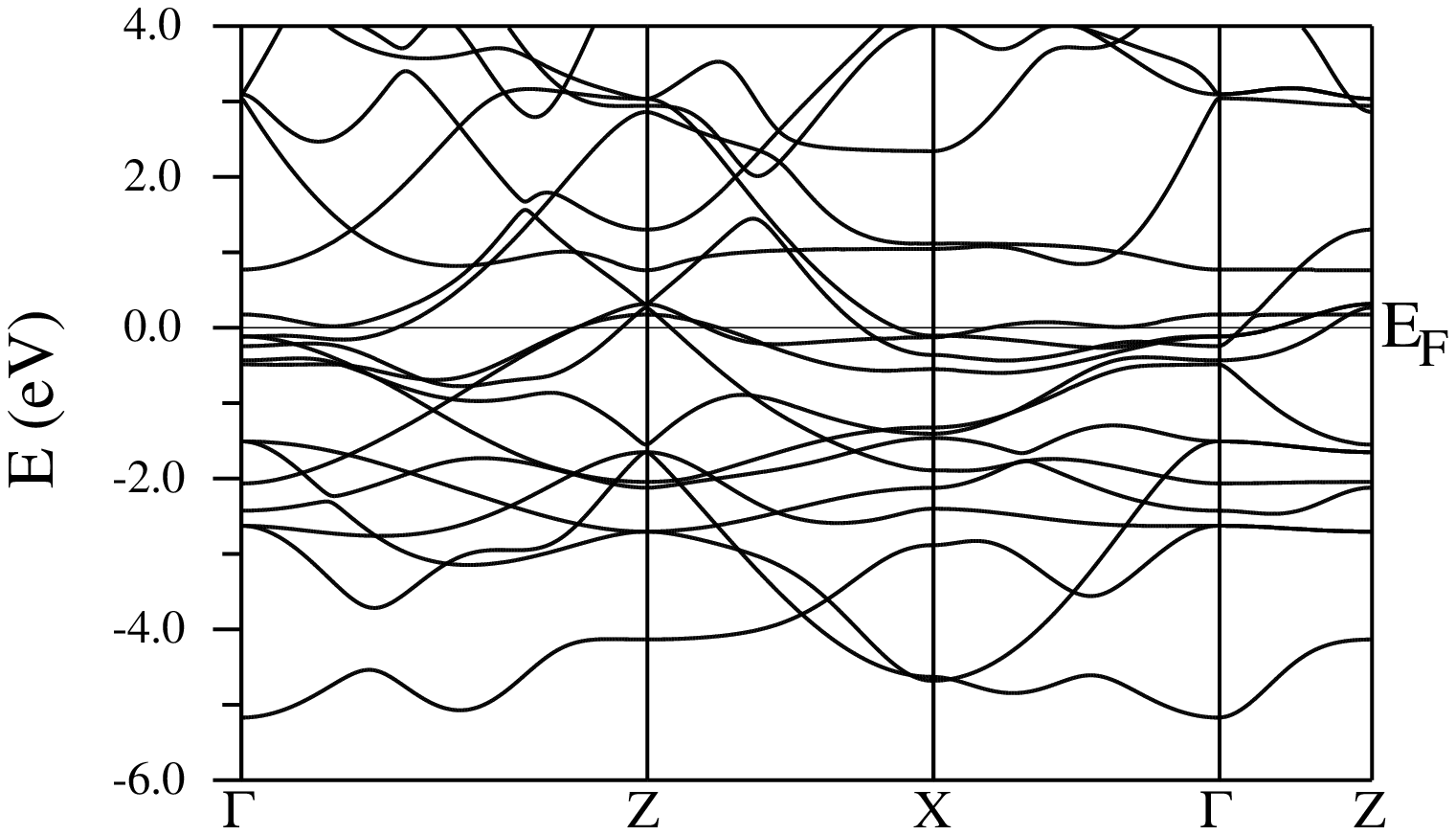}\\
  \includegraphics[width=\columnwidth]{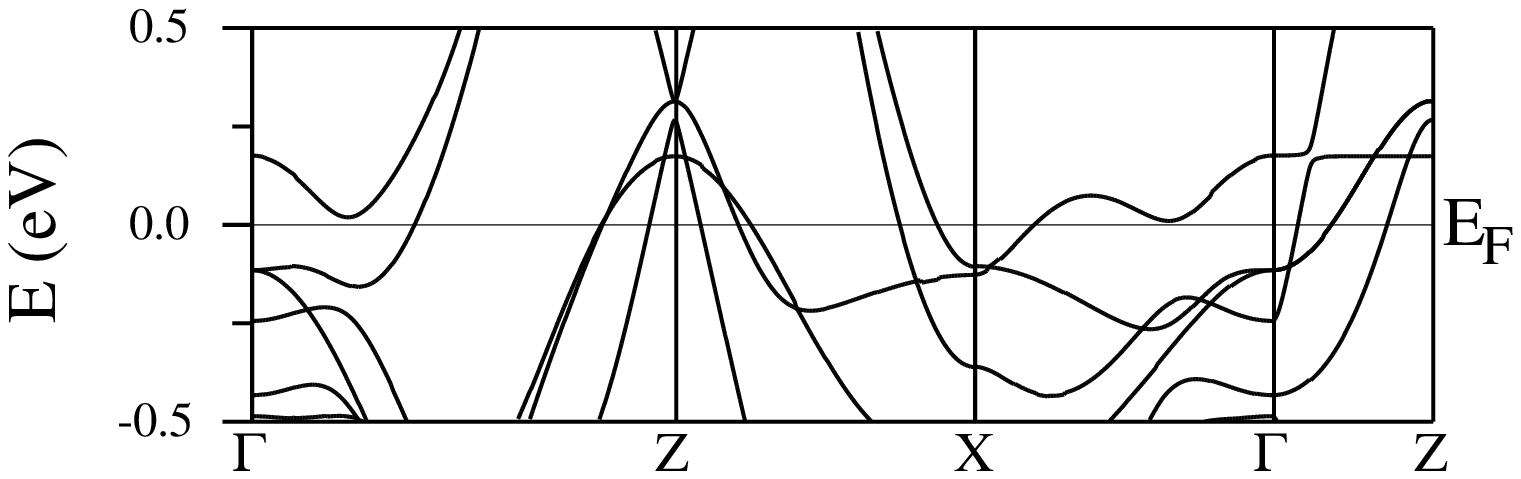}
  \caption{
  Top: LDA non-spin-polarized band structure of YFe$_2$Ge$_2$. Bottom:
  A blow-up of the band structure around Fermi level. The long
  $\Gamma$--$Z$ direction is from $(0,0,0)$ to $(1,0,0)$ and the
  short one is from $(0,0,0)$ to $(0, 0, 1/2)$. The $X$ point is
  $(1/2,1/2,0)$. The stacking of the Brillouin zone is such that
  $(1,0,0) = (0,0,1/2)$. See Fig.~1 of Ref.~\onlinecite{park10} for a
  particularly illuminating illustration of the reciprocal-space
  structure. 
    }
  \label{fig:yfg-bnd}
\end{figure}

\begin{figure} %% [tbp]
  \includegraphics[width=\columnwidth]{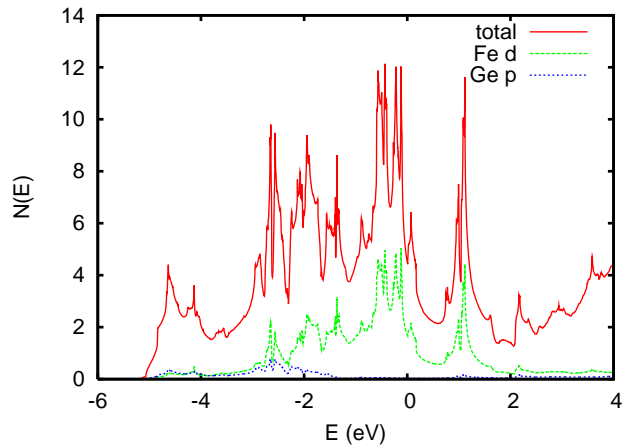}
  \caption{ (Color online) Electronic density of states
    non-spin-polarized of YFe$_2$Ge$_2$ and projections on to the LAPW
    spheres per formula unit both spin basis. }
  \label{fig:yfg-dos}
\end{figure}

The non-spin-polarized LDA band structure and density of states (DOS)
are shown in Figs.~\ref{fig:yfg-bnd} and \ref{fig:yfg-dos},
respectively. The lowest band that starts out from $\Gamma$ at $-$5.2
eV relative to the Fermi energy has Ge $4p_z$ character. There is only
one band with Ge $4p_z$ character below Fermi level, and there is
another band with this character above the Fermi level. This indicates
that the Ge ions make covalent bonds along the $c$ axis, which is not
surprising given the short Ge--Ge distance in that direction. The four
bands between $-$1.2 and $-$4.8 eV that start out from $\Gamma$ at
$-$1.5 and $-$2.6 eV have Ge $4p_x$ and $4p_y$ character. Rest of the
bands below the Fermi level have mostly Fe $3d$ character. 
Similar to the other iron-based superconductors,\cite{sing08a} there is
no gap-like structure among the Fe $3d$ bands splitting them into a
lower lying $e_g$ and higher lying $t_{2g}$ states. This shows that
Fe--Ge covalency is minimal and direct Fe--Fe interactions dominate.
Almost all of the Fe $4s$ and Y $4d$ and $5s$ character lie above the
Fermi level. This indicates a nominal occupation of Fe $3d^{6.5}$,
although the actual occupancy will be different because there is some
covalency of Fe $3d$ states with Y $4d$ and Ge $4p$ states.

The electronic states near the Fermi level come from Fe $3d$ derived
bands and show a rich structure. The electronic DOS at the Fermi level
is $N(E_F) = 4.50$ eV$^{-1}$ on a per formula unit both spin basis
corresponding to a calculated Sommerfeld coefficient of 10.63 mJ/mol
K$^2$. The Fermi level lies at the bottom of a valley with a large
peak due to bands of mostly $d_{xz}$ and $d_{yz}$ characters on the
left and a small peak due to a band of mostly $d_{xy}$ character on
the right. (The local coordinate system of the Fe site is rotated by
45$^\circ$ in the $xy$ plane with respect to the global cartesian axes
such that the Fe $d_{x^2-y^2}$ orbital points away from the Ge $p_x$
and $p_y$ orbitals.)  There is a pair of linearly dispersive band with
mostly $d_{xz}$ and $d_{yz}$ as well as noticeable Ge $p_z$ characters
either side of $Z$. If they are not gapped in the superconducting
state, they will provide the system with a massless excitation. In
addition to this pair of linearly dispersive bands, there is also a
very flat band near the Fermi level along $X$--$\Gamma$. This band has
an electron-like nature around $X$ and crosses the Fermi level close
to it. Along the $X$--$\Gamma$ direction, it reaches a maximum at 0.08
eV above the Fermi level, turns back down coming within 0.01 eV of
touching the Fermi level, and again moves away from the Fermi
level. It may be possible to access these band critical points that
have vanishing quasiparticle velocities via small perturbations due to
impurities, doping, or changes in structural parameters. The role of
such band critical points in quantum criticality has been emphasized
recently,\cite{neal11} and similar physics may be relevant in this
system.

\begin{figure}
  \includegraphics[width=0.8\columnwidth]{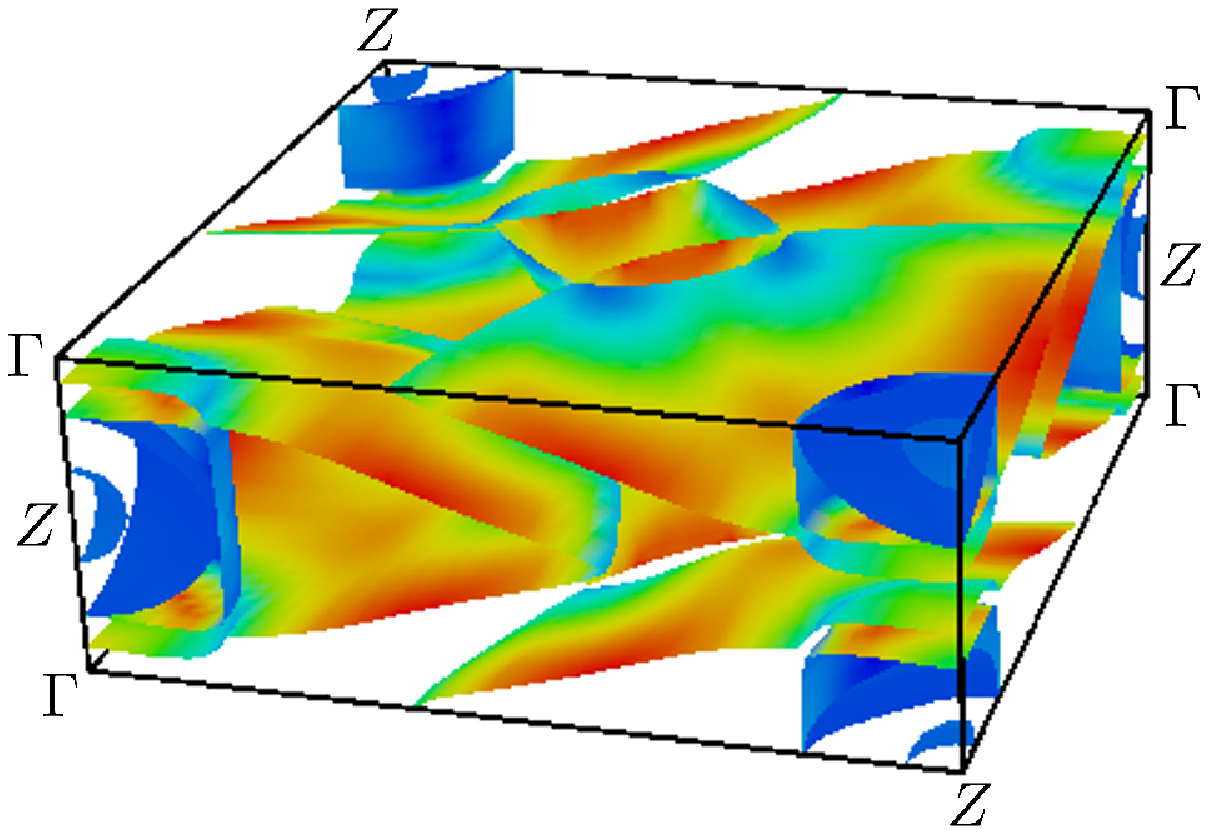}
  \includegraphics[width=0.8\columnwidth]{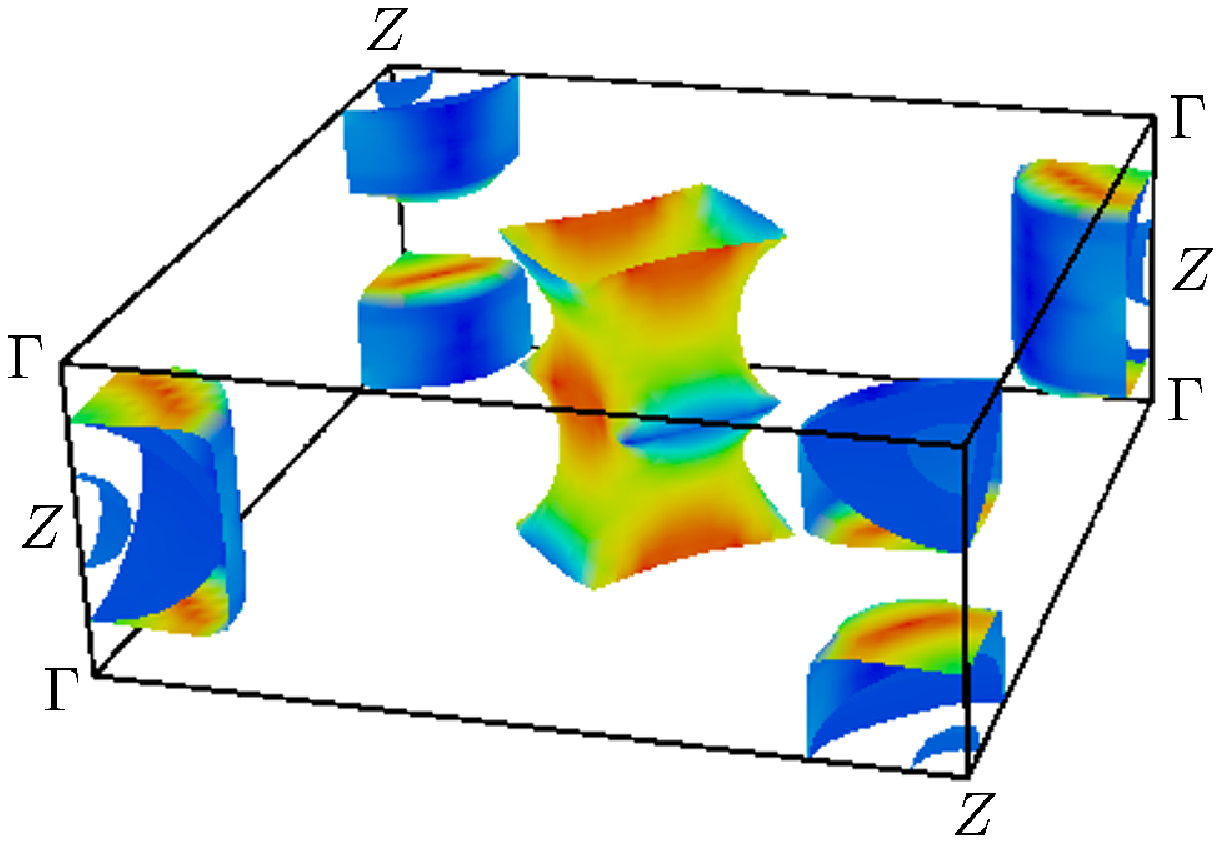}
  \caption{(Color online) Top: LDA Fermi surface of
    YFe$_2$Ge$_2$. Bottom: The Fermi surface without the large
    sheet. The shading is by velocity.}
  \label{fig:yfg-fs}
\end{figure}

The Fermi surface of this compound is shown in
Fig.~\ref{fig:yfg-fs}. There is an open very two dimensional
tetragonal electron cylinder around $X$. This has mostly $d_{xz}$ and
$d_{yz}$ character. There are four closed sheets around $Z$. One of
them is a large three dimensional sheet with the shape like the shell
of a clam with $d_{xz}$, $d_{yz}$, $d_{xy}$, and $d_{z^2}$
characters. There are two almost spherical hole sheets. These have
mostly $d_{xz}$ and $d_{yz}$ characters, with the smaller one also
containing noticeable Ge $p_z$ character. These two spherical sheets
are enclosed by a closed cylindrical hole sheet that has mostly
$d_{xy}$ character.

The cylindrical and larger spherical sheets centered around $Z$ touch
at isolated points. Otherwise, the Fermi surface is comprised of
disconnected sheets. If one considers the $\Gamma$--$Z$--$\Gamma$ path
along the $k_z$ direction, there is a series of box-shaped cylindrical
hole sheet that encloses the two spherical sheets. Although there are
no sections around $\Gamma$, these sheets around $Z$ enclose almost
two-third of the $\Gamma$--$Z$--$\Gamma$ path. Therefore, there is
likely to be substantial nesting between the sheets around $Z$ and $X$
that will lead to a peak in the susceptibility at the wave vector
$(1/2,1/2)$.

\begin{figure}
  \includegraphics[width=0.6\columnwidth]{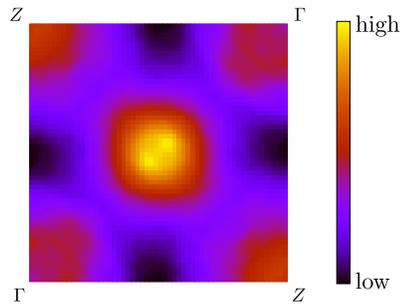}
  \caption{The real part of bare susceptibility calculated with the
    matrix element set to unity.}
  \label{fig:yfg-suscep}
\end{figure}

I have calculated the Lindhard susceptibility 
\[
\chi_0(q,\omega) = \sum_{k,m,n} |M_{k,k+q}^{m,n}|^2
\frac{f(\epsilon_k^m) - f(\epsilon_{k+q}^n)}{\epsilon_k^m -
  \epsilon_{k+q}^n - \omega - \imath \delta}
\]
at $\omega \to 0$ and $\delta \to 0$, where $\epsilon_k^m$ is the
energy of a band $m$ at wave vector $k$ and $f$ is the Fermi
distribution function. $M$ is the matrix element, which is set to
unity. 
%
%% This is not a drastic approximation as the largest contribution
%% to the peak in the susceptibility will come from the nesting between
%% the cylindrical sheets around $Z$ and $X$ that have similar orbital
%% character. 
%
The real part of the susceptibility is shown in
Fig.~\ref{fig:yfg-suscep}, and it shows peaks at $\Gamma$, $Z$, and
$X$ with the peak at $X$ having the highest magnitude. Note, however,
that the cylinders around $Z$ and $X$ have different characters, which
should reduce the peak $X$ and make it broader as well. The peak at
$\Gamma$ is equal to the DOS $N(E_F)$. The peak at $Z$ reflect the
nesting along the flat sections of the sheets along $(0,0,1/2)$
direction, while the peak at $X$ is due to the nesting between the
hole cylinder and spheres centered around $Z$ and the electron
cylinder centered around $X$.

The bare Lindhard susceptibility is further enhanced due to the RPA
interaction, and its real part is related to magnetism and
superconductivity. It is found experimentally that pure YFe$_2$Ge$_2$
does not order magnetically down to a temperature of 2 K although it
shows non-Fermi liquid behavior in the transport and heat capacity
measurements that is likely due to proximity to a magnetic quantum
critical point.\cite{zou13} As the temperature is lowered further,
superconductivity manifests in the resistivity measurements at
$T_c^\rho$ = 1.8 K and DC magnetization at $T_c^{\textrm{mag}}$ = 1.5
K. This superconductivity can be due spin fluctuations associated with
the peak in the susceptibility.\cite{berk66,fay80} The pairing
interaction has the form
\[
V(q=k-k') = - \frac{I^2(q) \chi_0(q)}{1 - I^2(q) \chi_0^2(q)}
\]
in the singlet channel and is repulsive. (In the triplet channel, the
interaction is attractive and also includes an angular factor.) Here
$I(q)$ is the Stoner parameter which microscopically derives from
Coulomb repulsion between electrons.

% The calculated susceptibility has a peak at $X$, and this favors a
% singlet $s_\pm$ superconductivity that changes sign between hole
% cylinder at $Z$ and electron cylinder at $X$ mediated by
% antiferromagnetic spin fluctuation with a wave vector $(1/2,1/2)$
% similar to that of the previously discovered iron-based
% superconductors \cite{mazi08}.

In the present case, the structure of the calculated susceptibility
leads to the off-diagonal component of the interaction matrix to have
a large negative value $-\lambda$ for the pairing between the hole
sheets at $Z$ and electron cylinder at $X$ in the singlet channel. The
diagonal component of the interaction matrix $\lambda_d$ pairing
interactions on the hole and electron sheets are small and
ferromagnetic. (For simplicity, I have assumed that the density of
states are same for the hole and electron sections.) The eigenvector
corresponding to the largest eigenvalue of this interaction matrix has
opposite signs between the hole sheets around $Z$ and electron
cylinder around $X$, and this is consistent with a singlet $s_\pm$
superconductivity with a wave vector $(1/2,1/2)$. This
superconductivity is similar to the previously discovered iron-based
superconductors.\cite{mazi08,kuro08}

The proposed superconducvity in YFe$_2$Ge$_2$ and the previously
discovered iron-based superconductor is similar, but the $T_c$ = 1.8 K
for YFe$_2$Ge$_2$ is much smaller than those reported for other
iron-based superconductors. One reason for this may be the smaller
nesting in this compound leading to a smaller peak in
susceptibility. The hole cylinder around $Z$ has mostly $d_{xy}$
character whereas the hole spheres around $Z$ and the electron
cylinder around $X$ have mostly $d_{xz}$ and $d_{yz}$ character. These
factors should lead to a slightly smaller and broader peak at $X$. I
note, however, that nesting in the other iron-based superconductors is
also not perfect\cite{mazi08} and the band characters between the
nested sheets also vary.\cite{kuro08}

Another reason for the smaller $T_c$ in YFe$_2$Ge$_2$ may be due to
the existence of competing magnetic fluctuations associated with the
proximity to quantum criticality. The DOS from non-spin-polarized
calculation is $N(E_F)$ = 1.125 eV$^{-1}$ per spin per Fe, which puts
this material on the verge of a ferromagnetic instability according to
the Stoner criterion. Ferromagnetism is pair-breaking for the singlet
pairing and will suppress the $T_c$ in this compound. Furthermore,
there is a peak in the susceptibility at $Z$ as well. The presence of
additional antiferromagnetic interactions might reduce the phase space
available for the spin fluctuation associated with the pairing channel
and may be pair-breaking as well.

\begin{table}[h!tbp]
  \caption{\label{tab:mag} The relative energies of various magnetic
    orderings and the moments within the Fe spheres. These are almost
    degenerate, indicating the proximity to quantum criticality is due
    to competing magnetic interactions.}
  \begin{ruledtabular}
    \begin{tabular}{lcc}
      & Energy (meV/Fe) & Moment ($\mu_B$/Fe) \\
      \hline
      NSP & 0 & 0 \\
      FM & $-$6.29 & 0.59 \\
      AFM (0,0,1/2) & $-$11.63 & 0.64 \\
      SDW (1/2,1/2,0) & $-$6.52 & 0.72
    \end{tabular}
  \end{ruledtabular}
\end{table}

I performed magnetic calculations with various orderings on $(1 \times
1 \times 2)$ and $(\sqrt{2} \times \sqrt{2} \times 2)$ supercells to
check the strength of competing magnetic interactions. The relative
energies and the Fe moments are summarized in Table \ref{tab:mag}. I
was able to stabilize various magnetic configurations, and their
energies are close to that of the non-spin-polarized
configuration. However, I was not able to stabilize the checkerboard
antiferromagnetic order in the $ab$ plane. When the magnetic order is
stabilized, the magnitude of the Fe moment is less than 1 $\mu_{B}$,
and the magnitudes vary between different orderings. This indicates
that the magnetism is of itinerant nature. It is worthwhile to note
that LDA calculations overestimate the magnetism in this compound as
it does not exhibit any magnetic order experimentally. This
disagreement between LDA and experiment is different from that for the
Mott insulating compounds where LDA in general underestimates the
magnetism.

Although this compound does not magnetically order experimentally, it
nonetheless shows proximity to magnetism. It is found that partial
substitution of Y by isovalent Lu causes the system to order
antiferromagnetically, with 81\% Lu substitution being the critical
composition.\cite{ran11} At substitution values below the critical
composition, the system shows non-Fermi liquid behavior in the heat
capacity and transport measurements.\cite{zou13} The unusually high
Sommerfeld coefficient of $\sim$90 mJ/mol K$^2$ at 2 K further
increases as the temperature is lowered and the resistivity varies as
$\rho \propto T^{3/2}$ up to around 10 K. This non-Fermi liquid
behavior and the large renormalization of the magnetic moments may
happen because there is a large phase for competing magnetic
tendencies in this compound. This is due to the
fluctuation-dissipation theorem, which relates the fluctuation of the
moment to the energy and momentum integrated imaginary part of the
susceptibility.\cite{mori85,solo95,ishi98,agua04,lars04} If the
quantum criticality is due to competing magnetic interactions, the
inelastic neutron scattering experiments, which measures the imaginary
part of the susceptibility, would exhibit the structure related to the
competing interactions. Therefore, even though this compound does not
show magnetic ordering, it would be useful to perform such experiments
and compare with the results presented here.

In any case, I indeed find that various magnetic orderings and the
non-spin-polarized configuration are close in energy (see Table
\ref{tab:mag}). The energy of the lowest magnetic configuration is
only 11.6 meV/Fe lower than the non-spin-polarized one, and the
energies of the different magnetic orderings are within 6 meV/Fe of
each other. As a comparison, the difference in energy between the
non-magnetic configuration and the most stable magnetic ordering in
BaFe$_2$As$_2$ is 92 meV/Fe, and the energy of the magnetic ordering
closest to the most stable one is higher by 51 meV.\cite{sing08b}
Signatures of quantum criticality has been reported for BaFe$_2$As$_2$
and related compounds.\cite{ning09,jian09,kasa10} YFe$_2$Ge$_2$
should show pronounced effects of proximity to quantum criticality as
the competition between magnetic interactions is even stronger.

In summary, I have discussed the superconductivity and quantum
criticality in YFe$_2$Ge$_2$ in terms of its electronic structure and
competing magnetic interactions. The electronic states near the Fermi
level are derived from Fe $3d$ bands and show a rich structure with
the presence of both linearly dispersive and heavy bands. The Fermi
surface consists of five sheets. There is an open rectangular electron
cylinder around $X$. A big sheet shaped like the shell of a clam
encloses a hole cylinder and two hole spheres around $Z$. There is a
peak in the bare susceptibility at $(1/2,1/2)$ due to nesting between
the hole sheets around $Z$ and the electron cylinder around $X$. I
propose that the superconductivity in YFe$_2$Ge$_2$ is due to
antiferromagnetic spin fluctuations associated with this peak. The
resulting superconducting state has a $s_\pm$ state similar to that of
previously discovered iron-based superconductors. I also find that
different magnetic configurations are close in energy, which suggests
the presence of competing magnetic interactions that are responsible
for the proximity to quantum criticality observed in this compound.

% \acknowledgments
I am grateful to Antoine Georges for helpful comments and
suggestions. This work was partially supported by a grant from Agence
Nationale de la Recherche (PNICTIDES).

% hole doping

\end{document}